\documentclass[11pt]{article} 

\usepackage{geometry}		

\usepackage[latin1]{inputenc}  
\usepackage{amsmath,amsfonts,amsthm,amscd,amssymb}  
\usepackage{dsfont}			
\usepackage{thmtools}		
\usepackage{nameref}

\theoremstyle{definition}

\declaretheoremstyle[
spaceabove=6pt, spacebelow=6pt,
headfont=\normalfont\bfseries,
notefont=\normalfont\bfseries, 
notebraces={}{},
bodyfont=\normalfont\itshape
]{Estilo1}

\declaretheorem[style=Estilo1,numbered=no,name=\!\!]{ThmName} 

\newcommand\ket[1]			{\left| #1 \right\rangle}
\newcommand\braket[2] 		{\left\langle #1 \mid #2 \right\rangle}
\newcommand\norm[1]			{\left|\left| #1 \right|\right|}

\newcommand{\Schrodinger}			{Schr\"o\-din\-ger}
\newcommand{\HH}{\mathcal{H}}
\newcommand{\E}{\mathcal{E}}
\newcommand{\M}{\mathcal{M}}
\newcommand{\U}{\mathcal{U}}
\newcommand{\OO}{\mathcal{O}}
\newcommand{\R}{\mathcal{R}}

\begin{document}

\title{Analysis of Wallace's Proof of the Born Rule in Everettian Quantum Mechanics II: Concepts and Axioms}

\author{Andr\'e L. G. Mandolesi \\ \emph{Departamento de Matem\'atica, Universidade Federal da Bahia} \\ \emph{Salvador-BA, Brazil} \\ \emph{E-mail:} \texttt{andre.mandolesi@ufba.br}}
\date{\today}  

\maketitle

\abstract{Having analyzed the formal aspects of Wallace's proof of the Born rule, we now discuss the concepts and axioms upon which it is built. Justification for most axioms is shown to be problematic, and at times contradictory. 
Some of the problems are caused by ambiguities in the concepts used. We conclude the axioms are not reasonable enough to be taken as mandates of rationality in Everettian Quantum Mechanics. This invalidates the interpretation of Wallace's result as meaning it would be rational for Everettian agents to decide using the Born rule.
}

\section{Introduction}

This paper is a continuation of a preceding one \cite{Mandolesi2017a}, in which Wallace's proof of the Born rule was reorganized and verified for formal errors. 
We proceed by checking his concepts for ambiguities, and questioning his justifications for the axioms.

To solve the measurement problem of the usual formulation of Quantum Mechanics, H.\,Everett\,III \cite{DeWitt1973,EverettIII1957} proposed in 1957 an alternative, known as the Many Worlds Interpretation. It rejects the Measurement Postulate (and the Born rule), and applies the rest of the usual formalism to all systems (even macroscopic ones), at all times (even during measurements). From this it should follow that the quantum state of an observer, after a measurement, is a superposition of distinct versions of himself, each correlated to one of the results and unaware of the others. 
Each component of such macroscopic superposition evolves independently, being called a world or branch.

But Everett's formulation has its own problems. The preferred basis one is whether there is a natural way to decompose a macroscopic quantum state into branches, and how similar to our (quasi-)classical world these are. D.\,Wallace \cite{Wallace2012} has proposed a solution, by combining the decoherent histories and Everettian formalisms. 

There is also a probability problem: how to make sense of quantum probabilities, in a deterministic theory in which all possible measurement results happen (even if in different worlds).
D.\,Deutsch \cite{Deutsch1999} has proposed using decision theory to show that, under Everettian conditions, it would be rational to decide on bets about the results of quantum measurements as if they were probabilistic and followed the Born rule.
Based on this idea, Wallace has produced a formal proof of the Born rule \cite{Wallace2010,Wallace2012}.
Having shown \cite{Mandolesi2017a} it to be, from a strictly formal point of view, mostly correct (some corrections were needed), in this article we continue our analysis  with a critique of the concepts and axioms upon which the proof is based. As we show, most are hard to justify from a more physical perspective.

A source of difficulties is that use of decision theory requires worlds where narratives with agents and bets make sense, so it depends on the preferred basis problem being solved. But Wallace's use of decoherence to solve it may implicitly rely on the Born rule, resulting in a circular argument \cite{Baker2007}.
Although he claims decoherence can be used without probabilities, we have shown \cite{Mandolesi2017} that, prior to the Born rule, one can not be sure the resulting branches would be sufficiently well behaved.

There are also ambiguities in concepts, like macrostate and reward, which are not  characterized in enough detail. This allows for incompatible interpretations, and Wallace seems to rely on different ones to justify various axioms.
Moreover, the formal statements of some axioms differ significantly from what he describes and justifies, while others include important details left unexplained.

Other authors \cite{Dizadji-Bahmani2013,Dizadji-Bahmani2013a,Jansson2016,Lewis2016,Mallah2008,Wilson2013} have criticized some of Wallace's axioms, but they have tended to focus on his informal presentation. Our analysis goes deeper into the formal details, contrasting them with their informal description. We also question ideas used to justify the axioms, identifying previously ignored problems in most of them. These may invalidate their proposed status as mandates of rationality, putting in doubt the meaning of Wallace's result.

In section \ref{sec:Preliminaries} we review the Everettian formalism, its problems, and proposed solutions, including results of our analysis of Wallace's solution to the preferred basis problem.
Section \ref{sec:Basic Concepts} presents the quantum decision problem and concepts used in the proof of the Born rule, with a critique of the ambiguities in their characterizations. In section \ref{sec:Axioms} Wallace's axioms are analyzed, and difficulties in their justifications are pointed out. Section \ref{sec:Conclusion} summarizes the problems that were identified.

\section{Preliminaries}\label{sec:Preliminaries}

\subsection{Everettian Quantum Mechanics}\label{sec:EQM}

The Many Worlds Interpretation, or Everettian Quantum Mechanics (EQM), is an attempt, proposed by H.\,Everett\,III \cite{DeWitt1973,EverettIII1957}, at solving problems of the usual Copenhagen interpretation of Quantum Mechanics (CQM), such as:
\begin{itemize}
\item \emph{measurement problem}: to characterize which quantum processes do not follow the deterministic \Schrodinger\ equation, constituting measurements; to explain the probabilistic nature of the results and the collapse of the wavefunction.
\item \emph{quantum-classical transition}: is quantum mechanics valid for all systems (micro- or macroscopic), with classical mechanics emerging as a particular case for large systems? If so, which mechanism prevents macroscopic quantum superpositions? 
\end{itemize}

In EQM the Measurement Postulate is rejected, and the rest of the quantum formalism is applied even to  macroscopic systems. Evolution is always deterministic, even in measurements, following \Schrodinger's equation.
It leads to macroscopic superpositions, but also explains why observers do not perceive them. If not for some unsolved problems, it might explain quantum measurements, and provide the connection between quantum and classical mechanics. 

A measurement, in this theory, is just quantum entanglement of the measuring device with whatever is being measured. More precisely, a \emph{measuring device} for a basis $\{\ket{i}\}$ of a system is any apparatus, in a quantum state $\ket{D}$, which interacts in such a way that, if the system is in state $\ket{i}$, the composite state evolves as\footnote{For simplicity, we assume the system remains in state $\ket{i}$, but this is not necessary.}
\begin{equation*}
\ket{i}\otimes\ket{D}\ \  \longmapsto\ \  \ket{i}\otimes\ket{D_i},
\end{equation*}
where $\ket{D_i}$ is a new state of the device, registering result $i$.
Linearity of Schr\"odinger's equation implies that, if the system is in a superposition 
\begin{equation}\label{eq:psi}
\ket{\psi}=\sum_i c_i\ket{i},
\end{equation}
the composite state evolves as
\begin{equation}\label{eq:MeasurementProcess}
\ket{\psi}\otimes\ket{D} = \left(\sum_i c_i\ket{i}\right)\otimes\ket{D} \ \  \longmapsto\ \   \sum_i c_i\ket{i}\otimes\ket{D_i}.
\end{equation}
This final state is to be accepted as an actual quantum superposition of macroscopic states. But it will not be perceived as such by an observer looking at the device, as, by the same argument, his state $\ket{O}$ will evolve into a superposition, according to
\begin{equation*}
\left(\sum_i c_i\ket{i}\otimes\ket{D_i}\right)\otimes\ket{O} \ \  \longmapsto\ \   \sum_i c_i\ket{i}\otimes\ket{D_i}\otimes\ket{O_i},
\end{equation*}
with $\ket{O_i}$ representing a state in which he saw result $i$. 
By linearity, each component $\ket{i}\otimes\ket{D_i}\otimes\ket{O_i}$ evolves independently, as if the others did not exist, as long as interference is negligible. In section \ref{sec:Preferred_Basis_Problem} we show why this assumption may be reasonable. 

Everett's interpretation of this final state is that the observer has actually split into different versions of himself, each seeing a result. Each version evolves as if the initial state had been $\ket{i}\otimes\ket{D}\otimes\ket{O}$, so he does not feel the splitting, nor the existence of his other versions.
Each component is called a \emph{world} or \emph{branch}, and this evolution of one world into a superposition of many is a \emph{branching process}. So in EQM all possible results of a measurement actually happen, but in different worlds. The observer in state $\ket{O_i}$ only thinks the system has collapsed into $\ket{i}$ because he does not see the whole picture, with all other results and versions of himself.

Problems plaguing CQM disappear in EQM, but new ones come along, such as the probability and preferred basis problems, described in the next sections.

\subsection{The Probability Problem}\label{sec:Probability_Problem}

When measuring \eqref{eq:psi}, in EQM, any result $i$ with $c_i \neq 0$ is obtained with certainty, even if only one version of the observer sees it. The \emph{probability problem} is to reconcile this with experiments, which suggest results are probabilistic and follow the Born rule:

\begin{ThmName}[Born Rule]
In CQM, when measuring \eqref{eq:psi} in an orthonormal basis $\{\ket{i}\}$, the probability of obtaining result $i$ is given by its \emph{Born weight} $w_i=\frac{|c_i|^2}{\|\psi\|^2}=\frac{|\braket{i}{\psi}|^2}{\braket{\psi}{\psi}}$.
\end{ThmName}

The problem has a qualitative aspect: how probabilities emerge in a deterministic theory. In classical mechanics processes can appear random due to ignorance of details, but in EQM we must explain the apparent randomness even if the quantum state and its evolution are perfectly known. 
Wallace \cite{Wallace2012} defends an operational and functional definition of probability, attained via decision theory and Bayesian inference. 

There is also the quantitative aspect of showing that probability values agree with the Born rule. A counting measure, based on the idea that a measurement with $n$ results produces $n$ branches, might seem most natural for EQM, in disagreement with that rule.
But Wallace argues there is no good way to count branches. In EQM measurements are no different than other quantum processes, so branchings can happen in all interactions, becoming a continuous and pervasive phenomenon. 
As most interactions involve few particles, many branches would be macroscopically similar, and a coarse-graining might reduce and stabilize their number. But such number would be somewhat arbitrary, depending on the chosen fineness of grain.

Everett \cite{EverettIII1957} proved that if a measure can be attributed to branches, and is preserved by further branchings, it equals the Born weights. And the total measure of  branches with results deviating from the Born rule tends to $0$, as the number of measurements tends to infinity. But for finite experiments this only means branches with frequencies deviating beyond a given error have small measure, which only makes them negligible if Born weights have a probabilistic interpretation, in a circular argument. A similar idea was proposed by Graham \cite{Graham1973}, with the same problem.

Gleason's theorem \cite{Gleason1957} also implies the Born rule, if the probability of a branch does not depend on what other branches there are in the decomposition basis. 
But until we know how probabilities emerge in EQM, we can not assume they satisfy such condition (of course, if they do not, it would mean EQM is incorrect).

Other attempts \cite{Albert1988,Buniy2006,Hanson2003,Zurek2005} have been made to solve the probability problem, without success. 
We focus on Deutsch and Wallace's use of decision theory.

Decision theory \cite{Karni2014,Kreps1988,Parmigiani.2009} shows that, in situations with probabilistic outcomes, rational (in an axiomatically defined sense) decisions follow a \emph{Principle of Maximization of Expected Utility}: 
a subjective utility value is attributed to each outcome, and choices with higher \emph{expected utility} (the probability-weighted average of utilities) are preferable.
When probabilities are unknown, L.\,Savage \cite{Savage1972} has shown that, with extra rational constraints, the principle remains valid, but with subjective probabilities. 

Deutsch \cite{Deutsch1999} proposed using this theory to show that, in an \emph{Everettian universe} (i.e. one governed by EQM), a rational decision maker should decide on bets about results of quantum experiments as if outcomes were probabilistic, with probabilities given by the Born weights. More precisely, he should follow the Principle of Maximization of Expected Utility, but with expected utilities redefined using Born weights in place of probabilities in the weighted average. 

Critics \cite{Barnum2000,Hemmo2006,Mallah2008,Price2006} contested his assumptions, proposed reasonable decision strategies violating his result, or questioned its meaning. In response, Wallace has made increasingly sophisticated attempts \cite{Wallace2003,Wallace2007,Wallace2010,Wallace2012} to clarify Deutsch's idea.
Although many authors \cite{Assis2011,Carroll.2014,Polley2001,Read2018,Saunders2010a,Sebens2016,Wilson2013} consider this approach promising, critics remain unconvinced \cite{Albert2010,Dawid2014,Finkelstein2009,Jansson2016,Kent2010,Maudlin2014,Price2010}, proposing counterexamples or questioning Wallace's axioms, though most have focused on his informal presentation.

In \cite{Mandolesi2017a} we have analyzed Wallace's proof \cite{Wallace2010,Wallace2012} from a formal perspective, showing it to be mostly correct (there are some problems but they seem fixable). In this article, however, we show that, from a more physical point of view, his concepts and axioms face serious problems, possibly invalidating his result.

\subsection{The Preferred Basis Problem}\label{sec:Preferred_Basis_Problem}

The \emph{preferred basis problem} is how to find a natural way to decompose a macroscopic quantum state into branches behaving like the classical reality we observe (even if not all of them, and not all the time).
We focus on Wallace's approach \cite{Wallace2012}, based on decoherent histories.

\subsubsection{Decoherence}\label{sec:Decoherence}

\emph{Decoherence} \cite{Joos2003,Schlosshauer2007,Zurek2002} is a process by which an open quantum system loses some quantum characteristics, as it interacts and gets entangled with its environment. 
Models show some \emph{pointer states} are more robust with respect to such interaction, having a stronger tendency to remain disentangled. 
For example, in the quantum brownian motion, considered a good paradigm for a macroscopic system scattering particles from the environment, they are minimum-uncertainty Gaussian packets (coherent states) $\ket{x,p}$, forming an overcomplete set of vectors. 

As the environment is differently affected by distinct pointer states, and such differences spread across its many degrees of freedom, it rapidly evolves into (almost) orthogonal states. Off-diagonal elements (\emph{coherences}) of the reduced density matrix of the system, with respect to the pointer states, decay extremely fast. This (almost) eliminates interference between such states, and we say the system has \emph{decohered}. 
Close pointer states take longer to decohere, but this problem can be reduced by coarse graining, as we discuss in section \ref{sec:Decoherent Histories}. As the reduced density matrix becomes (nearly) diagonal, it formally resembles a classical probabilistic mixture, and this is seen as an important step in obtaining classicality out of quantum systems.

Evolution of an open quantum system is usually studied using a master equation for its reduced density matrix. But it lacks unitarity, which is central to Wallace's arguments. So he does not take such approach, turning instead to the consistent/decoherent histories formalisms, which apply to closed quantum systems.

\subsubsection{Consistent Histories}\label{sec:Consistent Histories}

The \emph{Consistent Histories} formalism \cite{Griffiths1984,Griffiths2002,Omnes1988,Omnes1999} identifies conditions allowing us to assign classical probabilities to sets of alternative histories, conceived as sequences of events or propositions about a quantum system.

Its point of view is opposite to the Everettian one: quantum evolution is always stochastic, and quantum states are only tools for calculating probabilities in a set of possible evolutions, only one of which actually happens. Still, parts of the formalism adapt well to the Everettian setting, if appropriately reinterpreted. We present a simplified version\footnote{Reflecting its stochastic approach, Consistent Histories is usually presented using density operators.}, assuming a normalized pure initial state $\psi_0$ at time $t_0$. 

A \emph{quantum sample space} is an orthogonal projective decomposition of Hilbert space, i.e. a family $\{P_\alpha\}$ of orthogonal projection operators with $P_{\alpha} P_{\beta}=\delta_{\alpha\beta}P_{\alpha}$ and $\sum_{\alpha} P_{\alpha}=\mathds{1}$, representing an exhaustive set of mutually exclusive events or propositions about the system. 
A \emph{history space} is a sequence $\{P_{\alpha_1}(t_1)\}, \ldots, \{P_{\alpha_n}(t_n)\}$ of quantum sample spaces, at times $t_0<t_1<\ldots<t_n$ (the Heisenberg picture is used, with
$ P_{\alpha}(t)=U(t,t_0)^{-1}P_{\alpha}U(t,t_0)$,
where $U(t,t_0)$ is the time evolution operator given by Schrodinger's equation).
A \emph{history} $\alpha=(\alpha_1,\ldots,\alpha_k,\ldots,\alpha_n)$ is a sequence of events, specifying one $P_{\alpha_k}(t_k)$ from each sample space. 

In each history space, satisfying a consistency condition described below, only one history happens, with probability postulated by
\begin{equation}\label{eq:probability_history}
p_\alpha = \norm{\psi_\alpha}^2, 
\end{equation}
where the \emph{branch state vector} $\psi_\alpha=P_{\alpha_n}(t_n) \cdots P_{\alpha_1}(t_1) \psi_0$ is to be considered as just a mathematical tool to calculate probabilities.

A history space $\{\bar{P}_{\bar{\alpha}_k}(t_k)\}$ is a \emph{coarse-graining} or \emph{coarsening} of $\{P_{\alpha_k}(t_k)\}$ if each $\bar{P}_{\bar{\alpha}_k}(t_k)$ is a sum of $P_{\alpha_k}(t_k)$'s. Conversely, $\{P_{\alpha_k}(t_k)\}$ is a \emph{fine-graining} or \emph{refinement} of $\{\bar{P}_{\bar{\alpha}_k}(t_k)\}$. 
To fit this in the formalism, history spaces are bundled into a \emph{history algebra}, specified by  taking, at each $t_k$, a \emph{quantum event algebra}, which is a Boolean algebra\footnote{With operations $P_1\wedge P_2=P_1P_2$, $P_1\vee P_2=P_1+P_2-P_1P_2$, and $\neg P=\mathds{1}-P$.} of orthogonal projectors.
The Boolean condition implies projectors commute, so when their ranges have  intersection $\{0\}$ they are mutually orthogonal. 

To get a history algebra, we can start with a history space and take all coarsenings. The original projectors are minimal elements (\emph{atoms}) generating the event algebras. Not all history algebras are formed this way, and some have no atoms. An algebra is \emph{atomic} if it is generated by atoms. 
Wallace uses the term in such sense, but defines it in terms of countable generators \cite[p.95]{Wallace2012}. This is incorrect, as any countable algebra is generated by the set of all its elements, even if it has no minimal ones.

Given a coarsening $\{\bar{P}_{\bar{\alpha}_k}(t_k)\}$ of $\{P_{\alpha_k}(t_k)\}$, each coarser history $\bar{\alpha}$ can be seen as a family of finer ones, with $\alpha\in\bar{\alpha}$ if, for each $k$, the range of $\bar{P}_{\bar{\alpha}_k}(t_k)$ contains that of $P_{\alpha_k}(t_k)$. 
The consistent histories formalism requires the $p_\alpha$'s to behave as classical probabilities, so they must be additive, $p_{\bar{\alpha}} = \sum_{\alpha\in\bar{\alpha}} p_\alpha$. 
Hence there must be no interference between histories, and $\langle \psi_\alpha | \psi_\beta \rangle=0$ if $\alpha\neq \beta$. 
History spaces satisfying this are called \emph{consistent}\footnote{The terminology varies. Wallace uses consistency for additivity of probabilities, and decoherence for non-interference, which gets mixed with the related, but not equivalent, concept from section \ref{sec:Decoherence}.} (relative to $\psi_0$), and are the only ones allowed in the formalism. 
Two consistent history spaces can be \emph{incompatible}, in the sense that they can not be combined into a single consistent one. The formalism allows either one to be used in describing a quantum process, but not both at once.

A history space has a \emph{branching structure} (relative to $\psi_0$) if histories do not merge after diverging, i.e. if $\alpha$ and $\beta$ satisfy $\alpha_i\neq\beta_i$ and $\alpha_j=\beta_j$ for some $i<j$, one of them has zero probability. Consistency and branching are related as follows \cite{Griffiths1993,Wallace2012}:

\begin{ThmName}[Branching-Consistency Theorem]\label{Branching-Consistency Theorem}\footnote{Branching-Decoherence Theorem, in Wallace's terminology.}
Any history space with a branching structure is consistent, and the converse holds for some consistent refinement of it.
\end{ThmName}

 The idea \cite[p.429]{Wallace2012} for the converse is that, since by consistency the $\psi_\alpha$'s are orthogonal, a branching refinement can be constructed from projectors $\frac{1}{p_\alpha}|\psi_\alpha\rangle \langle \psi_\alpha|$. Wallace claims \cite[p.96]{Wallace2012} this result extends to history algebras, which is wrong, as these projectors might not be in the original algebra.

\subsubsection{Decoherent Histories}\label{sec:Decoherent Histories}

Consistent Histories admits histories that are far from classical.  
The \emph{Decoherent Histories} formalism, developed by Gell-Mann and Hartle \cite{Gell-Mann1990,Gell-Mann1993}, combines Consistent Histories and decoherence to get special history spaces with a more classical behavior. 
Besides (approximate) consistency\footnote{In Gell-Mann and Hartle's terminology, medium decoherence.}, it also seeks \emph{quasi-classicality}, meaning histories approximately follow classical equations of motion, interrupted at times (as in quantum measurements) by some quantum behavior. 

In this formalism, the space of relevant variables of the system of interest is partitioned in cells $\Sigma_\alpha$, large in comparison with the coherence lenght (below which decoherence is not effective), yet small enough for the required precision. This determines a quantum sample space given by operators $P_\alpha \otimes \mathds{1}_{\text{env}}$, where
\begin{equation}\label{eq:P_alpha}
P_\alpha=\int_{\Sigma_\alpha} \mathrm{d}{x} |{x}\rangle\langle{x}|, 
\end{equation}
and $\mathds{1}_{\text{env}}$ is the identity operator for the environment (taken to include the irrelevant variables of the system). 
A history $\alpha$ specifies, at a sequence of times $t_k$, in which $\Sigma_{\alpha_k}$ the system is\footnote{If necessary, different families of cells can be used at each time.}. 
Coarsenings and refinements are obtained varying cell size.

For large enough cells, states in the range of different $P_\alpha$'s are distinct enough to quickly get entangled to (almost) orthogonal states of the environment. Orthogonality tends to subsist, as the environment evolves, for its many degrees of freedom \emph{keep records} of the history. For example, particles scattered by the system in different directions, at distinct $\Sigma_\alpha$'s, tend to remain in (almost) orthogonal states, and affect other degrees of freedom in distinct ways. 
Erasing from the environment all traces of the history is impossible in practice. 
Of course, as different states of the environment evolve, they can spontaneously develop similar components, with blurred records, but their Born weights would be negligibly small. 

As records ensure (almost) consistency, (almost) additive probabilities can be assigned to histories, as in Consistent Histories.
Some interference can remain, but it is negligible if histories are coarse enough. Deviations in the additivity of probabilities should be irrelevant, as long as they are too small to be detected experimentally.

The \nameref{Branching-Consistency Theorem} ensures some refinement has a branching structure.
But it will not be in terms of projectors $P_{\tilde{\alpha}}\otimes \mathds{1}_{\text{env}}$ into smaller cells $\Sigma_{\tilde{\alpha}}$, requiring projectors to separate different records in the environment. And the branching structure will be approximate: there will actually be a mesh of tiny probability branches connecting and interfering (negligibly) with all main branches.

Models show that, if histories are coarse enough, and the system has enough inertia to resist noise from the environment, histories with non-negligible probabilities will (approximately) follow classical equations of motion, with a stochastic force.

So (almost) consistent history spaces seem to emerge naturally in Decoherent Histories, and their non-negligible histories are quasi-classical. 
However, we can still have incompatible descriptions of the evolution, as shown by Dowker and Kent \cite{Dowker1996}.

\subsubsection{Non-Probabilistic Decoherent Histories}\label{sec:Non-Probabilistic Decoherent Histories}

Here we review the conclusions of our analysis of Wallace's proposed adaptation  \cite{Wallace2012} of Decoherent Histories to solve the preferred basis problem, referring to  \cite{Mandolesi2017} for a more detailed discussion.

The trouble with his solution is that, as EQM is not stochastic and all histories happen, probabilities can not be postulated as in \eqref{eq:probability_history}, they must be obtained proving the Born rule. But Wallace's proof of such rule depends on branches provided by Decoherent Histories, forming a circular argument \cite{Baker2007}.

To avoid this, his proof of the Born rule would have to be compatible with \emph{Non-Probabilistic Decoherent Histories} \cite{Mandolesi2017}, meaning the non-probabilistic parts of the Decoherent Histories formalism, reinterpreted from an Everettian perspective. In this handicapped formalism, approximations indicated in section \ref{sec:Decoherent Histories}, usually justified on the basis of negligible probabilities, must be reexamined.

Wallace has suggested \cite[p.253]{Wallace2012} non-probabilistic reasons for them, but they do not seem valid. His arguments rely on the assumption that EQM is a reasonable physical theory, in which small perturbations in a quantum state do not drastically change the emergent structure of worlds. But the trouble with EQM is precisely that we are not sure it makes reasonable predictions. 

One may argue no physical model describes reality with absolute precision, so it is only fair that we accept Wallace's approximations. However, while in most theories approximations correspond just to slight deviations in the values of some quantities, in Wallace's case the state predicted by his model and the approximate one he wishes to use (with tiny components discarded) are literally worlds apart.

If we are to take EQM and Wallace's use of Decoherent Histories seriously, any cell $\Sigma_\alpha$ in which the wavefunction does not vanish completely corresponds to some branch. Whether such world is irrelevant when its Born weight is small is a question that must wait until the Born rule is proven, unless another reason is found. Even with the Born rule, we would have to question what does it mean to consider a world (and its inhabitants) irrelevant just because its Born weight can be treated as if it represented a low probability, when the theory tells us such world definitely exists. 

A possible solution might be to consider only \emph{causal histories} \cite{Mandolesi2017} (branches with Born weight large enough to resist interference and sustain causal relations) as corresponding to actual worlds. But that formalism is still in its early stages of development, and it may face a serious problem: if only branches with Born weights above some threshold represent real worlds, what happens to a world just above such cutoff once a quantum measurement splits it into smaller branches? Is our world headed towards extinction as we perform more and more quantum experiments?

In any case, if we accept that small components can be neglected prior to any probabilistic interpretation, we may as well forget Wallace's proof of the Born rule and accept Everett's simpler one.

Until the formalism of Non-Probabilistic Decoherent Histories is complemented by the Born rule, we must admit it could have strange features, as discussed in \cite{Mandolesi2017}:

\begin{enumerate}
\item Worlds represented by small amplitude branches must be considered as real and relevant as large ones. This causes \emph{branch discontinuities}: arbitrarily small changes in a quantum state can introduce all kinds of different worlds. 

\item As wavefunctions tend to spread, generating tiny components far away, (nearly) all macroscopic histories happen and are equally relevant, no matter how erratic. The branching structure disappears into the mesh described in section \ref{sec:Decoherent Histories}, as no branches stand out anymore for being more probable.

\item An alternative view is that there is no real macroscopic history, as all macroscopic states tend to exist at all times. For one such state to be absent the wavefunction must totally vanish in the corresponding cell, which only happens exceptionally.

\item Branches can exhibit macroscopic quantum jumps all the time, and records are unreliable, compromising any macroscopic sense of causality. 
Even branches allowing a meaningful macroscopic narrative keep sprouting weird subbranches. 

\item Tiny amplitude branches suffer significant interference from much larger ones.

\item A branch can follow classical equations without resembling our world, as it might not even have atoms, whose stability depends on quantum probabilities.

\item Decoherence does not happen in terms of sharp boundary cells, as pointer states $\ket{\pi_x}$ tend to be peaked at a point $x$, but have small tails across all space. Usually it does not matter, as the tails represent negligible probabilities. Without such interpretation, \eqref{eq:P_alpha}  should be replaced by the following projectors, which can give more stable branches, but are only almost orthogonal to each other:
\begin{equation}\label{eq:pointer projector}
P_\alpha=\int_{\Sigma_\alpha} \mathrm{d}{x} |{\pi_x}\rangle\langle{\pi_x}|.
\end{equation}

\item Orthogonality might not mean in EQM the same as in CQM, as the measurement process \eqref{eq:MeasurementProcess} can occur in an almost orthogonal basis (perturbing its states to preserve unitarity). So orthogonal states might not be mutually exclusive when measured. An almost orthogonal set of branches can be linearly dependent, so, even allowing for coarsenings or refinements, decompositions might be non unique. And two decompositions of a state can involve quite different sets of branches.
\end{enumerate}

These weird results suggest that Wallace's proposed mix of  EQM and Decoherent Histories (prior to a probabilistic interpretation) might not be viable.
If the Born rule can be proven, despite these problems, they might not matter anymore, affecting only very unlikely branches. But in proving such rule they must be taken into account.

\section{Quantum Decision Problem and Related Concepts}\label{sec:Basic Concepts}

We present Wallace's quantum decision problem, discussing the concepts involved and the framework in which it is set up.  The characterizations of some concepts are shown to be vague, with ambiguities that will affect the justifications for the axioms. Also, some concepts are supposed to emerge from the solution of the preferred basis problem, which, as seen in section \ref{sec:Non-Probabilistic Decoherent Histories}, is problematic prior to the Born rule.

\subsection{A Framework for Quantum Decision Theory} 

To analyze Wallace's concepts and axioms, we must first recognize the framework in which they are set. For this, it is important to have in mind that EQM is not (yet) just Universal Unitary Quantum Mechanics (UUQM), by which we mean the unitary part of the usual quantum formalism (i.e. discarding the Measurement Postulate), applied to all systems, at all times. Rather, EQM can be roughly described as UUQM plus the claim that, 
with the proper interpretation, (branchwise) classical mechanics and the usual quantum results (including the Born rule) emerge from it somehow.

As Wallace aims to justify such claim, he can not take EQM as fully valid from the start. He can not start from UUQM either, as such formalism is still too raw to provide the classical ingredients he needs: a decision theoretic proof requires a framework in which narratives with agents, bets and payoffs make sense. 
His starting point seems to be what we call Almost Everettian Quantum Mechanics (AEQM): UUQM plus a hypothesis that a branching structure of worlds, reasonably similar to ours, emerge from it, with the Born rule being the only missing ingredient for full EQM (and assuming that lack of such rule does not impair the rest of the picture).

This may seem valid if one believes Wallace's solution of the preferred basis problem bridges the gap from UUQM to AEQM. But, as seen in section \ref{sec:Non-Probabilistic Decoherent Histories}, without the Born rule his solution can lead to something quite different. Validity of some of his concepts and axioms will depend on a better justification for AEQM.

Also, the hypothesis of AEQM should be properly detailed. Are branches mutually orthogonal? Is decomposition unique? Is interference negligible? Are all branches similar to our world? How similar? 
If intelligent life exists, does it reason in ways we can comprehend, or has it evolved to use another logic, better adapted to a branching universe? Is some kind of rationality even possible? 

 Wallace's assumptions regarding these questions, though not always clearly stated, are implicit in his concepts and axioms, whose validity depends on the answers being properly justified.
 Even if at times he admits an Everettian universe might not be as expected, for the most part he seems to believe his solution of the preferred basis problem provides branches similar enough to our world to have all required properties. 
One must however set aside any unwillingness to admit consequences of the formalism other than the desired ones, and consider how the possibilities of section \ref{sec:Non-Probabilistic Decoherent Histories} impact his concepts and axioms.

\subsection{Quantum Decision Problem}\label{sec:Quantum Decision Problem}

Wallace describes \cite[p.163]{Wallace2012} a \emph{quantum decision problem} as one where, in an Everettian universe, a system prepared in some state is to be measured. Bets are available, giving in each branch a payoff depending on the corresponding result. And an Everettian agent (someone who knows EQM correctly describes his universe, and knows the Born weights of that state in the measurement basis) must decide which bets he prefers. Formally, the problem is specified by:
\begin{itemize}
\item A separable Hilbert space $\HH$.
\item An \emph{event space} $\E$, which is a complete Boolean algebra of subspaces of $\HH$ (with $\HH\in\E$), under the operations of conjunction $\wedge$ (intersection), disjunction $\vee$ (closure of the span of the union), and orthogonal complement $\perp$. A \emph{partition} of $E\in\E$ is a set of mutually orthogonal events whose disjunction is $E$.
\item A set of \emph{macrostates} $\M$, which is a subset of $\E$ such that any $E\in\E$ has a partition in macrostates.
\item A set of \emph{rewards} $\R$, which is a partition of $\HH$.
\item For each $E\in\E$, a set $\U_E$ of unitary operators from $E$ into $\HH$, the \emph{acts available at $E$}. The following conditions are required for any $E,F\in\E$, where $U\lvert_E$ is the restriction of $U$ to $E$, $\OO_U$ is the smallest event containing the range of $U$, and $\mathds{1}_E$ is the identity act at $E$:
\begin{itemize}
\item \emph{Restriction}: If $F\subset E$ and  $U\in\U_E$ then  $U\lvert_F\in\U_F$. 
\item \emph{Composition}: If $U\in\U_E$ and $V\in\U_{\OO_U}$ then $VU\in\U_E$.
\item \emph{Indolence}: If \ $\U_E\neq\emptyset$ then $\mathds{1}_E\in\U_E$.
\item \emph{Continuation}: If $U\in\U_E$ then $\U_{\OO_U}\neq\emptyset$.
\item \emph{Irreversibility}: If $U\in\U_{E\vee F}$ then $\OO_{U\lvert_E}\wedge \OO_{U\lvert_F}=\emptyset$.\footnote{As noted in \cite{Mandolesi2017a}, it should be $\{0\}$ instead of $\emptyset$, and there is a missing hypothesis, that $E\wedge F=\{0\}$.}
\end{itemize}
\end{itemize}

A solution is given by a \emph{preference order} $\succ^\psi$ on $\U_M$, for each state $\psi$ of each $M\in\M$, satisfying Wallace's axioms.

Justifications for the axioms depend on how the above concepts 
are interpreted. Wallace hints at how he conceives of them, but important details are left out, allowing for conflicting interpretations. 
As he adopts a framework provided by the decoherent histories formalism, $\HH$ should to be the Hilbert space of the total system of interest (agent, measuring device, payoffs, etc.), plus its environment.

\subsection{Events}\label{sec:Events}

Intuitively, an \emph{event} $E\in\E$ is the subspace spanned by all states satisfying some proposition. 
Operators $\wedge$, $\vee$, $\perp$ play in Quantum Logic \cite{Birkhoff1936,Engesser2009} roles similar to the connectives AND, OR, NOT of Classical Logic. 

An important difference between quantum and classical (Boolean) logics is that the distributive law fails. 
By requiring $\E$ to be a Boolean algebra, Wallace is imposing a classical structure on it, in the spirit of Consistent Histories. In fact, it is equivalent to the quantum event algebras. 
Wallace uses the same one at all times, as for him events represent \emph{macroproperties}, emerging naturally in EQM via decoherence.
The Boolean condition implies, via distributivity, the following one:
\begin{ThmName}[Orthogonality Condition]\label{ax:OrthCond}
Conjunction of two events $E$ and $F$ is zero if, and only if, they are orthogonal, i.e.
$ E\wedge F=\{0\}\ \Leftrightarrow\ E\perp F.$
\end{ThmName}

Events $E$ and $F$ are \emph{disjoint} or \emph{mutually exclusive} if, whenever a state of $E$ is tested for the condition of $F$, the result is false, and vice versa. Classically, the same concept means no state satisfies both conditions, and is equivalent to $E\cap F=\emptyset$. In CQM this is replaced by $E\perp F$, since measuring the property of $F$ in a state nonorthogonal to it can result true (in EQM, it definitely results true in some branch). The \nameref{ax:OrthCond} makes quantum disjointness similar to the classical one, by forbidding the use of sets of events which are not ``classical enough''.

For example, we can not have in the same quantum decision problem an event characterized by an electron having spin up and another in which it has spin in some non-vertical direction. But we might have an event in which its measurement resulted up and another in which it was measured in the non-vertical direction, as long as the states of the measuring devices have enough differences to be orthogonal.

To partition an event is to decompose it in mutually exclusive subevents.
Given partitions $\{E_i\}$ and $\{F_j\}$ of the same event, $\{F_j\}$ is a \emph{refinement} of $\{E_i\}$, and $\{E_i\}$ is a \emph{coarsening} of $\{F_j\}$, if each $E_i$ admits a partition in terms of $F_j$'s. With the \nameref{ax:OrthCond}, two partitions of an event always have a common refinement.

As stated in section \ref{sec:Non-Probabilistic Decoherent Histories}, orthogonality might not play in EQM its usual role. If almost orthogonal measurements are possible, $E\perp F$ no longer ensures mutual exclusivity (in fact, there might be no mutually exclusive events): measuring a state of $E$ in an appropriate almost orthogonal basis might produce a branch in $F$, which, despite its tiny Born weight, can not be neglected. Likewise, use of $\perp$ as a negation operator might not be valid.

\subsection{Macrostates}\label{sec:Macrostates}

Wallace states \cite[p.164]{Wallace2012} that ``the choice of macrostates is largely fixed by decoherence, although the precise fineness of grain of the decomposition is underspecified''. It must not be too coarse, so ``an agent can be assumed not to care exactly what the microstate is within a given macrostate''. Also, ``an agent can have no practical control as to what state she gets, within a particular macrostate, on familiar statistical-mechanics and decoherence grounds'' \cite[p.170]{Wallace2012}. 

From this, and the way Wallace uses macrostates, we infer they  are to determine the state of the system of interest at a macroscopic level, playing the role of classical states.
He seems to have in mind the ranges of operators $P_\alpha \otimes \mathds{1}_{\text{env}}$ used in decoherent histories. Different families of cells $\Sigma_\alpha$ allow for refinements or coarsenings. Cells should be large enough for decoherence, yet small enough that the agent is indifferent to points in the same cell.
States in the same $M\in\M$ should differ, for the system of interest, only on microscopic details. There can be macroscopic differences in the environment, but we are to assume they are beyond the agent's control and he does not care about them.

However, the \nameref{ax:Irrev} condition (see section \ref{sec:Axioms}) may require  macrostates to be further refined by partitioning the Hilbert space of the environment according to all possible history records. So each macrostate should specify not only the macroscopic state of the system of interest, but also the history that led to such state. 

Given a partition of $E\in\E$ into macrostates $M_i\in\M$, we say $\psi\in E$ has a \emph{branch decomposition} with \emph{branches} $\psi_i\in M_i$ if $\psi=\sum_i\psi_i$.
As such partitions always exit, events are disjunctions of macrostates satisfying some common condition, and $\psi\in E$ if it is decomposable in branches having such condition.
 
Note that most states are not in any event other than $\HH$, as any such event is a proper subspace. Another way to look at this is that a typical wavefunction has a tail spreading through all cells. In CQM this tail represents tiny probabilities and can be neglected, but if EQM is taken seriously, and macrostates are defined as above, it means a typical state has branches in all possible macrostates. So, as stated in section \ref{sec:Non-Probabilistic Decoherent Histories}, the idea of a macroscopic history might be physically meaningless in this formalism, as all macrostates tend to be present at all times.

If the $P_\alpha$'s are as in \eqref{eq:P_alpha}, the \nameref{ax:OrthCond} holds, but macrostates evolve erratically, splitting all the time into all possible branches, as seen in section \ref{sec:Non-Probabilistic Decoherent Histories}. This makes them bad models for classical states, and Wallace's axioms hard to justify.

With the more realist projectors \eqref{eq:pointer projector}, we might get more stability, but at the cost of orthogonality, and the Boolean condition, being satisfied only approximately, with all consequences entailed by almost orthogonality.
For example, while with the \nameref{ax:OrthCond} two branch decompositions of $\psi$ have a common refinement,
with almost orthogonality decomposition may be non unique, even up to refinements. In fact, even a single branch might be decomposable into a family including physically quite distinct branches, with wavefunctions peaked at very distant cells.

The formal definition of $\M$ is quite flexible: it can be any subset of $\E$ which generates it via disjunctions. We can even have $\M=\E$, or take it to be any set of orthogonal subspaces whose disjunction is $\HH$, with $\E$ being the Boolean algebra it generates. Wallace mentions these as legitimate possibilities \cite[p.176]{Wallace2012}, but they allow for extreme cases, such as 1-dimensional macrostates, or a macrostate consisting of the whole Hilbert space. Such artificial examples do not fit with his informal description of macrostates, or how he uses them in justifying his axioms.

He also describes \cite[p.165]{Wallace2012} an event space $\E$ formed by functions supported in open sets of $\mathds{R}^N$. He says it can not be generated by macrostates, relating it to atomless history spaces \cite[p.95]{Wallace2012}, to show why he would not ask that $\E$ be constructible from $\M$. This is confusing, as his definitions do require it. 
The problem here seems to be the confusion, seen in section \ref{sec:Consistent Histories}, between atoms and generators.
In any case, the example is wrong: as the topology of $\mathds{R}^N$ is second countable, $\E$ is generated by a countable subset of events, which can be taken as $\M$.

\subsection{Rewards}\label{sec:Rewards}

Rewards ``represent payoffs an agent could get'' \cite[p.175]{Wallace2012}, and are
``a coarse-graining of the macrostate subspaces\ldots such that an agent's only preference is to which reward subspace she is in'' \cite[p. 165]{Wallace2012}.

In classical decision theory the terms \emph{reward} or \emph{payoff} are used in a broad sense, referring not only to cash or prizes awarded but also to expenses or losses the agent may incur, and can include other factors deemed relevant, such as any enjoyment, effort or time expenditure involved in the betting process. 

Wallace uses these terms somewhat loosely, which causes some confusion.
By his formal definition of $\R$, rewards are events, hence subspaces.
At times he talks about them as if they were tangible prizes, like cash (e.g. in his justification for the \nameref{ax:ReAv} axiom \cite[p.167]{Wallace2012}), but in general he seems to use the term in its broader sense, to represent anything that might affect the agent's preferences. 

To avoid such mix-up, we adopt the following terminology:
\begin{description}
\item[\emph{prize:}] cash, trophies or other tangible elements awarded in a bet. The term can also be used for negative elements, like the price paid to participate in a bet. But it must not include anything that might be affected by extraneous quantum measurements or by erasing information about the bet;
\item[\emph{payoff:}] any factor affecting the agent's preferences. Besides prizes, time, enjoyment, and so on, it can even include branchings and information erasures used in the betting process, if it turns out he might care about them (as we discuss later);
\item[\emph{reward:}] the element of $\R$ representing a payoff. As payoffs include not just prizes but preference factors in general, some confusion might have been avoided if Wallace had called it \emph{preference subspace} instead of reward.
\end{description}

Classically it is not hard to identify which factors the agent might care about, if we have a good intuition about the situation.
But intuition might not be so reliable in the Everettian case.
As we discuss later, it is not even clear whether the agent should care about extra branchings.

Even if we knew all factors which might affect preferences, Wallace does not detail how $\R$ is to be determined by them. 
The \nameref{ax:BrIndif} axiom provides a clue (if $\psi$ and $\phi$ are in the same reward, with $\psi$ in a macrostate, the agent does not care about going from $\psi$ to $\phi$), but it is not enough to characterize $\R$.
Taking equivalence classes of states according to preferences (i.e. $\psi$ and $\phi$ are in the same reward iff the agent is indifferent about being taken from one to the other, or perhaps from any third state to either of them) does not work, as rewards would not be mutually orthogonal. It seems we must first form equivalence classes of macrostates, and then take the disjunction of each class to be a reward.

Wallace's choice of defining rewards as subspaces (a payoff could as well have been represented by the set of macrostates corresponding to it, instead of their disjunction) embeds an important assumption about the agent's preferences: if two states are in the same reward, so the agent is indifferent between them, he must also be indifferent between any of them and a superposition of both. 
The variety rule, proposed by Elga \cite[p.193]{Wallace2012}, is an example that a superposition of states might be preferable to each. Maudlin \cite{Maudlin2014} discusses this question in more detail.

The description of $\R$ as a coarse-graining of $\M$ seems to indicate any $M\in\M$ is in some $r\in\R$. And, indeed, a macrostate intercepting more than one reward would contradict Wallace's description of these concepts. But his use of $M\wedge r$ in the \nameref{ax:MacIndif} axiom (see section \ref{sec:Axioms}) suggests otherwise.

\subsection{Acts}\label{sec:Acts}

Intuitivelly, an act $U$ is to represent any action of interest, like measuring a quantum state, placing a bet, receiving a payoff, or compositions of them. Availability of an act depends on the event, e.g. the act of deciding a bet is only available at events where it has been placed.

In EQM, even macroscopic evolutions are described by unitary operators. 
But this assumes closed systems, so each act $U$ describes not only actions performed by the agent and related to the decision problem, but also everything else happening in the environment (ultimately, the whole Universe). This can include extraneous facts, like a friend getting sick, which could alter his preference for one $U$ over another. Perhaps one could select for comparison only acts which do not differ significantly in their effects on the environment, or equivalence classes of acts, but it is not clear how to do so in practice.

A branched state has many versions of the agent, each acting on his own branch $M_i$. In EQM their individual acts $U_i\in\U_{M_i}$ are restrictions of an $U\in\U_{\vee_i M_i}$. Wallace says they form a \emph{compatible act function} (we just say they are \emph{compatible}). Being the $M_i$'s mutually orthogonal, so must be their images $U(M_i)$. 
So acts in different branches are linked, threatening the autonomy of the agent's versions.
But if evolution of the $M_i$'s preserves macroscopic differences, and the environment keeps (approximate) records of the macroscopic history, the $U_i$'s tend to have almost orthogonal images, and small adjustments might be enough to get perfect orthogonality. But we should worry whether such adjustments can be made without altering significant aspects of the acts. For example, with Non-Probabilistic Decoherent Histories they could cause the creation or elimination of worlds which, despite their small Born weights, can not be neglected.

The formal definition of $\U_E$ suggests there can be events with no available acts, i.e. $\U_E=\emptyset$. Wallace provides no explanation for this, and such events play no role in his proof. In fact, it seems there should be no such events.
As evolution via \Schrodinger's equation does not preserve compact support, a wavefunction supported in some limited region immediately develops a tail spreading through all space, i.e. it branches into all possible macrostates. If each branch evolves independently, we should not expect the wavefuntion to retreat into a limited region. So in general an act $U$ should have $\OO_U=\HH$. By Continuation, $\U_\HH\neq\emptyset$, so by Restriction $\U_E\neq\emptyset$ for any $E\in\E$. Hence Indolence and Continuation could be replaced by a simpler condition, that $\mathds{1}_E\in\U_E$ for all $E\in\E$.

Conditions of the form $\U_E\neq\emptyset$ do appear in some axioms (although originally concealed in Wallace's definition of \emph{available set of events} \cite[p.176]{Wallace2012}), so they might somehow be relevant for their validity. Again, Wallace gives no explanation. As we show in the discussion of \nameref{ax:ReAv}, it seems their role is just to provide an escape from cases in which the axiom is clearly not valid.

\section{Axioms}\label{sec:Axioms}

Despite using a more explicit notation and terminology, we state Wallace's axioms as in \cite{Wallace2012}, although some corrections are needed, as shown in \cite{Mandolesi2017a}.
We show the arguments used to justify them are flawed, and sometimes rely on contradictory ideas. Also, their formal statements include important details left unexplained, and at times differ considerably from their informal descriptions. Moreover, some axioms are compromised by the conceptual problems discussed above.

\subsection{Richness Axioms}\label{sec:Richness Axioms}

These are conditions concerning the acts available to the agent.

\begin{ThmName}[Irreversibility]\label{ax:Irrev}
Let $E,F\in\E$ with $E\wedge F=\{0\}$. If $U\in\U_{E\vee F}$ then $\OO_{U\lvert_E}\wedge \OO_{U\lvert_F}=\{0\}$.
\end{ThmName}

Wallace does not include this as an axiom, leaving it as a condition in the definition of $\U_E$, but it is important enough to deserve close attention. Here we have stated it with the corrections mentioned before.

The idea of \nameref{ax:Irrev} is that of a branching structure: evolution of distinct branches can not generate a common one, so they do not interfere. 
In \cite[p.88]{Wallace2012}, Wallace invokes the usual arguments of decoherent histories, that preservation of records in the environment keeps branches from interfering. But, as seen in section \ref{sec:Non-Probabilistic Decoherent Histories}, with Non-Probabilistic Decoherent Histories records are not reliable, and there is always some interference, even if it is small.

He also says an agent in a branch can not cause others to merge \cite[p.164]{Wallace2012}. 
But even if he can not do it on purpose, violations of irreversibility can happen naturally, e.g. when a wavepacket crosses the boundary of adjacent cells. 
And, again, with Non-Probabilistic Decoherent Histories small violations happen all the time, and can have a significant impact on tiny (but as relevant as any) branches. 

Later Wallace discusses irreversibility at lenght \cite[ch.9]{Wallace2012}, tracing parallels with the Second Law of Thermodinamics. But at that point he assumes the Born rule as proven \cite[p.333]{Wallace2012}, so his conclusions there can not justify this condition. 
Also, note that most physical laws are reversible, the Measurement Postulate and the Second Law being notable exceptions. Both are poorly understood, and might be linked somehow, so that, without the first, it is possible the other might also be invalidated in EQM. In any case, if macrostates refer only to the macroscopic state of the system of interest, the Second Law would not prevent branches from merging, it would only enact a cost for this, in terms of an increase in the environmental entropy.

He also appeals \cite[p.334]{Wallace2012} to the \nameref{Branching-Consistency Theorem}. But, as seen in section \ref{sec:Consistent Histories}, it does not apply to history algebras, so the branching structure might not be in terms of macrostates, unless these separate history records in the environment. 
Anyway, decoherence gives only almost consistency, so the branching structure would be approximate, with small violations of irreversibility.

If violations are small enough they might not compromise Wallace's proof, as long as the \nameref{ax:StaSup} axiom remains valid with approximate equalities. But, as we discuss later, this might not hold.

\begin{ThmName}[Reward Availability]\label{ax:ReAv}
Given a set $\{M_i\}$ of mutually orthogonal macrostates, with $\U_{\vee_iM_i}\neq\emptyset$, and for each $i$ a reward subspace $r_i\in\R$, there is  $U\in\U_{\vee_iM_i}$ such that $\OO_{U\lvert_{M_i}}\subset r_i$ for all $i$.
\end{ThmName}

Wallace's justification for the availability of such \emph{reward act} is that ``envelopes of cash can always be given to people'' \cite[p.167]{Wallace2012}. But his reward subspaces were not characterized in terms of prizes, like cash, but in terms of the agent's preferences. 
Unless we are sure to know all an Everettian agent might care about, how can we say there is always a physically realizable action leading to any such subspace?

For example, as we discuss later, we do not agree with Wallace's assumption that the agent is indifferent to branching acts. If he cares about them, they might take him from one reward subspace into another. A reward act sending him back to the original reward might require undoing the branching, in a violation of \nameref{ax:Irrev}.

Given a partition of $\HH$ in macrostates $\{M_i\}$, and $r\in\R$, the axiom would require an unitary map $U:\HH\rightarrow \HH$ with $U(\HH)\subset r$, which is impossible if $r\neq\HH$. The only way out is if $\U_\HH=\emptyset$, but, as discussed in section \ref{sec:Acts}, this does not seem to be valid. 
Moreover, there seems to be no reason for including the condition $\U_{\vee_iM_i}\neq\emptyset$ in the axiom, or for admitting events with no available acts, other than to create a loophole in an attempt to avoid such contradictions.

Even if $E$ is a proper subspace of $\HH$, it may be hard to obtain $\OO_{U\lvert_E}\subset r$. As seen in section \ref{sec:Acts}, as a wavefunction evolves it tends to spread through all space, so most acts have $\OO_{U}=\HH$. Even if some have $\OO_{U}\subset r$, arguments of imprecision, as those Wallace uses for \nameref{ax:PrCont}, imply that in practice we can never be sure to get them. 
What seems feasible is to obtain an act producing only very small components outside $r$, but until we have the Born rule these can not be neglected.
Saying our experience shows cash can be given without generating extraneous branches would be meaningless, as our universe might not be governed by EQM.
One might also argue that this is just an idealization, with the agent contemplating such acts in his decision as if they could be executed with precision. In our discussion of \nameref{ax:SolCont} we explain why idealizations like this can lead to wrong conclusions. We note that this problem will also affect \nameref{ax:BrAv} and \nameref{ax:Eras}.

\begin{ThmName}[Branching Availability]\label{ax:BrAv}
Let there be given:
\begin{itemize}
\item a set $\{M_i\}$ of mutually orthogonal macrostates, with $\U_{\vee_iM_i}\neq\emptyset$, and with each $M_i$ in some $r_i\in\R$;
\item for each $i$, a nonzero $\psi_i\in M_i$ and a set $\{p_{ij}\}$ of positive numbers with $\sum_j p_{ij}=1$.
\end{itemize}
Then there is $U\in\U_{\vee_iM_i}$ such that, for each $i$, 
\begin{itemize}
\item $\OO_{U\lvert_{M_i}}\subset r_i$;
\item there is a partition $\OO_{U\lvert_{M_i}}=\vee\!_j\, N_{ij}$ with $N_{ij}\in\M$ and  $\|\Pi_{N_{ij}} U\psi_i\|^2/\|\psi_i\|^2=p_{ij}$.
\end{itemize}
\end{ThmName}

Such \emph{branching act} is explained by the possibility of preparing and measuring an arbitrary quantum state. Again, Wallace does not discuss the hypothesis $\U_{\vee_iM_i}\neq\emptyset$.

Prizes are not altered in the act, so if reward subspaces were determined by them this would explain why $\OO_{U\lvert_{M_i}}\subset r_i$. But as $\R$ was characterized in terms of the agent's preferences, this condition means he can not care about any aspects of the act. 
This anticipates the \nameref{ax:BrIndif} axiom, and will be discussed with it. 

\begin{ThmName}[Erasure]\label{ax:Eras}
Let there be given:
\begin{itemize}
\item two countable sets $\{M_i\}$ and $\{N_i\}$ of mutually orthogonal macrostates, such that $\U_{\vee_i M_i}\neq\emptyset$, $\U_{\vee_i N_i}\neq\emptyset$ and, for each $i$, we have $M_i, N_i \subset r_i$ for some $r_i\in\R$;
\item for each $i$, nonzero states $\psi_i\in M_i$ and $\phi_i\in N_i$.
\end{itemize} 
Then there are $U\in\U_{\vee_i M_i}$ and $V\in\U_{\vee_i N_i}$ such that, for each $i$, $\OO_{U\lvert_{M_i}}, \OO_{V\lvert_{N_i}}\subset r_i$ and $U\psi_i =V\phi_i$.
\end{ThmName}

As noted in \cite{Mandolesi2017a}, this axiom has a missing hypothesis, that $\|\psi_i\|=\|\phi_i\|$, which is needed by unitarity, since Wallace does not use normalized states.

To explain such \emph{erasures}, Wallace says ``an agent can just forget any facts about his situation that don't concern things he cares about (i.e. by definition: that don't concern where in the reward subspace he is)'' \cite[p.167]{Wallace2012}.
He suggests these acts take the state of each branch into an \emph{erasure subspace}, available at any reward, ``whose states correspond to the agent throwing the preparation system away after receiving the payoff but without recording the actual result of the measurement''. 
And, as he ``lacks the fine control to know which act he is performing, all erasures should be counted as available if any are. It follows that, since for any two such agents all erasures are available, in particular there will be two erasures available satisfying the axiom''.

If $\psi_i$ and $\phi_i$ are in the same reward $r_i$, they are equivalent in all features relevant to the agent, and erasures must eliminate the remaining differences. The axiom says the final states must be  identical in all microscopic details, not only macroscopically similar. So it entails erasing all information about any differences from the environment and the agent's memory, which are part of the quantum system.
This contradicts ideas used to justify \nameref{ax:Irrev}, which assumed information is preserved in environmental records. Unless the agent has the unrealistic capability of eliminating all history traces  from the environment, the sets of states to which $\psi_i$ and $\phi_i$ can be taken, even if by different acts, should be almost orthogonal, invalidating this axiom.

Even if erasures can be made to work at a pair of branches, the axiom requires them to be done at many pairs at once. 
As seen in section \ref{sec:Acts},  acts on different branches are not completely autonomous, but small adjustments might make them compatible. However, they might perturb the condition $U\psi_i=V\phi_i$. 

Let us try to detail Wallace's arguments to see how they fail.
He seems to assume the agent can go from any state of $r_i$ into an erasure subspace $e_{r_i}\subset r_i$, and that all ways to do this (from any given macrostate) are so similar he can not distinguish them (perhaps Wallace thinks $e_{r_i}$ is a single macrostate). From this he seems to conclude the agent can always reach any state of $e_{r_i}$, making it easy to obtain $U\psi_i=V\phi_i$.

But let $M_1$ and $M_2$ be orthogonal macrostates in the same reward $r$. By Wallace's arguments, an agent at $\psi_1\in M_1$ can perform an act $U_1$ such that $U_1\psi_1\in e_{r}$, while another version of himself at $\psi_2\in M_2$ performs an $U_2$ with $U_2\psi_2\in e_{r}$, with both acts being compatible, i.e. restrictions of an $U\in\U_{M_1\vee M_2}$. As \nameref{ax:Irrev} gives $\OO_{U_1}\wedge\OO_{U_2}=\{0\}$, $e_{r}$ has states with enough differences to be in disjoint events. So there is no reason to expect that all acts leading into $e_r$ be indistinguishable, even if they start at the same macrostate.

Still, let $M$ and $N$ be orthogonal macrostates in $r$, $\psi\in M$ and $\phi\in N$, and suppose there are erasures $U\in\U_M$ and $V\in\U_N$ such that $U\psi=V\phi$. If acts on distinct branches can indeed always be made compatible, at a superposition $\psi+\phi$ the agent's versions could perform $U$ and $V$ at the same time (perhaps with small adjustments), so they would be restrictions of a $W\in\U_{M\vee N}$. But then $W\psi=W\phi$ (or at least they are close states), violating unitarity.

As before, the condition $\OO_{U\lvert_{M_i}}, \OO_{V\lvert_{N_i}}\subset r_i$ means the agent can not care about the processes involved in such acts, and it will be discussed with \nameref{ax:BrIndif}.

\begin{ThmName}[Problem Continuity]\label{ax:PrCont}
	For any $E\in\E$ with $\U_E\neq\emptyset$, $\U_E$ is an open subset (in operator norm topology) of the set of unitary operators from $E$ to $\HH$.
\end{ThmName}

This axiom will be discussed with \nameref{ax:SolCont}. We just note here that the condition $\U_E\neq\emptyset$ is irrelevant, as in topology the empty set is open by definition.

\subsection{Preference Axioms}\label{sec:Preference Axioms}

These are conditions the agent's preference orders $\succ^\psi$ must satisfy. 
Wallace labels them \emph{rationality axioms}, and calls a decision strategy \emph{rational} if it satisfies them. This allows him \cite[pp.189--197]{Wallace2012} to dismiss counterexamples as irrational if they contradict his axioms (as if such contradictions could not possibly be resolved in detriment of the axioms). Or to claim (\cite[p.237]{Wallace2010},  \cite[p.195]{Wallace2012}) it is possible to decide rationally in an Everettian universe, since the strategy given by the Born rule satisfies all axioms. 
To avoid such mix-up of an arbitrary label with the normal use of ``rational'' as meaning reasonable or intelligent, we prefer to call them \emph{preference axioms}, at least until their reasonability (or not) can be established. 

Equating a set of axioms with rationality is usual in classical decision theory. But of course this is not arbitrary, there are reasons why the classical axioms are accepted as mandates of rationality: they may seem intuitively reasonable, or it can be shown that following them tends to give good results more frequently (e.g. via the law of large numbers), while violations can lead to bad consequences.

In the Everettian case, can we confirm the rationality of Wallace's axioms by showing the resulting decision strategy (to follow the Born rule) gives the best possible results? Hemmo and Pitowsky \cite{Hemmo2006} claim the strategy is irrational, as some versions of the agent get bad results. Wallace \cite[p.195]{Wallace2012} says this problem is not particular to EQM: in probabilistic settings some people get unlucky. 
Indeed, bad branches only invalidate the Born rule if another strategy produces less of them, but this is meaningless without a way to quantify branches. On the other hand, it also means we can not say the Born rule maximizes good branches (of course, the measure given by the rule can not be used to corroborate it).
And even if there is a way to measure branches, other criteria for evaluating results are conceivable in a branching Universe (e.g. the distributive-justice and variety rules \cite[p.192]{Wallace2012}).

We can not trust intuition either, as an Everettian universe could be quite different from ours. So any arguments for the rationality of Wallace's axioms had better be quite convincing, leaving no room for different possibilities.
He has presented many, and at first sight they may seem reasonable, albeit for wrong reasons: resemblance of what is rational in our universe, or an implicit expectation that the Born rule be valid. 
As we show, there are several reasons to consider them invalid.

\begin{ThmName}[Ordering]\label{ax:Ord}
For each $M\in\M$ and $\psi\in M$, $\succ^\psi$ is a total order on $\U_M$.
\end{ThmName}

This axiom comes from classical decision theory, and seems reasonable enough.

\begin{ThmName}[State Supervenience]\label{ax:StaSup}
	Let $M, M'\in\M$, $\psi\in M$, $\psi'\in M'$, $U, V\in\U_{M}$ and $U', V'\in\U_{M'}$. If $U\psi=U'\psi'$ and $V\psi=V'\psi'$  then $U\succ^{\psi} V \Leftrightarrow U'\succ^{\psi'} V'$.
\end{ThmName}

This means preferences can not depend on the initial states or the betting process, only on the final states. 
In classical decision theory, a bet is defined by its payoffs and their probabilities, with no description of the process or initial state. This simplification is usually harmless, as intuition allows us to adjust the model, including in the description of payoffs factors like time spent or effort. But in the Everettian case it is not so easy, since we do not know what all relevant factors are. 

A possible argument (not presented by Wallace) could be that the final state embeds information, in environmental records and the agent's memories, about the act and initial state. 
But this fails with erasures, which are supposed to destroy such information, and in Non-Probabilistic Decoherent Histories, where records are unreliable. In Wallace's proof, this axiom is used precisely after erasures.

Wallace does not explain why the initial status of the agent (at the moment of decision) can not have any bearing on his preferences.
To explain why all acts connecting the same initial and final states are equally preferable, he says \cite[p.171]{Wallace2012} they differ only on states the agent should not care about, for not being the actual one. But he notes this could be a limitation of his formalism, which uses single unitary operators, not detailing the evolution process. The histories formalism would allow more flexibility, but the agent would have to be indifferent ``to the size of the temporal gaps between history projectors'', and as ``we consider sequences of decisions made only over very short periods of time'', this ``entails that acts can be represented by single unitary transformations''.

This is not convincing. Bets are not so quick that nothing relevant can happen in the process. In Wallace's proof they have intermediary steps, with branching acts and erasures, which might be pertinent, as we discuss later. And lack of a natural temporal gap does not keep us from placing quantum sample spaces at important steps. 

As discussed, for Wallace's proof to work with small violations of \nameref{ax:Irrev} this axiom would have to be valid even if the final states were only close to each other, instead of equal. But, as seen in section \ref{sec:Non-Probabilistic Decoherent Histories}, even small differences in the states can represent entirely different worlds, which, despite their tiny Born weights, can not be neglected and might affect preferences.

For further discussion of this axiom, see \cite{Jansson2016}.

\begin{ThmName}[Diachronic Consistency]\label{ax:DiacCons}
	Let $M\in\M$, $\psi\in M$, $U\in\U_{M}$ and $V, V'\in\U_{\OO_U}$. Given a partition $\OO_U=\vee_i M_i$ with $M_i\in\M$, let  $\phi_i=\Pi_{M_i}U\psi$. Then:
	\begin{itemize}
		\item if $V\lvert_{M_i}\succcurlyeq^{\phi_i} V'\lvert_{M_i}$ for all $i$ with $M_i$ not null\,\footnote{$M$ is \emph{null for $\psi$ and $U$} iff, whenever acts $V_1$ and $V_2$ are identical on $M^\perp$, $V_1U\sim_\psi V_2U$, i.e. the agent at $\psi$ is indifferent to what happens in $M$ after $U$. The reason should be that $U\psi$ has no branches in $M$, but there are problems with this concept \cite{Mandolesi2017a}.} for $\psi$ and $U$, then $VU\succcurlyeq^{\psi} V'U$; 
		\item if, in addition, $V\lvert_{M_i}\succ^{\phi_i} V'\lvert_{M_i}$ for at least one such $i$, then $VU\succ^{\psi} V'U$.
	\end{itemize} 
\end{ThmName}

Wallace says this axiom ``rules out the possibility of a conflict of interest between an agent and his future selves'' \cite[p.168]{Wallace2012}. But, as shown in \cite{Mandolesi2017a}, such temporal aspect is only part of the axiom, and it actually follows from \nameref{ax:StaSup}. If $\phi=U\psi$ has a single branch, that axiom gives $V\succ^{\phi} V' \Leftrightarrow VU\succ^\psi V'U$. And it can be extended to branched states, by defining $V\succ^{\phi} V'$ to mean all versions of the agent at $\phi$ prefer the restriction of $V$ to their branch over that of $V'$.

The really new idea introduced by this axiom is Branch Independence \cite{Mandolesi2017a}: preference between acts, at a branched state, is independent of branches in which their restrictions are equivalent for the agent, being determined only by those where he has a strict preference. Wallace does not discuss this idea, which is an Everettian version of the classically controversial Independence or Sure-Thing Principle \cite{Parmigiani.2009}.
To see the problem with it, consider the following bets:
\begin{itemize}
	\item[$A$:] an electron spin, with equal Born weights for up and down, is measured, with the agent receiving \$1,000 in up branches, and paying \$100 in down ones. 
	\item[$B$:] an electron spin, prepared in a state like the one above, is measured, with the agent receiving \$1,000 in up branches, and nothing in down ones. 
	\item[$C$:] bet $B$ is performed, and in the down branches it is followed by $A$.
\end{itemize}

In $A$ the agent has more to gain than to lose (assuming he can afford to pay \$100), so we can assume he might prefer it over doing nothing, i.e. $A\succ \mathds{1}$. Note that we do not need the Born rule to justify this, we can simply use the symmetry of the state. \nameref{ax:DiacCons} then implies $C\succ B$.

But $C$ gives a set of branches where he gets \$1,000 and one where he pays \$100, while $B$ produces a set where he receives \$1,000 and another where he loses nothing.
To prefer $C$ over $B$ he must think the \$1,000 set is in some sense larger or more relevant in $C$ than in $B$, to compensate for the branches in which he has a loss. So we need him to believe there is a physically meaningful measure, even if unknown, which is increased by the new \$1,000 branches resulting from $A$. But there seems to be no reason for such belief to be mandatory in EQM, if branch counting is impossible and we can not assume he knows or accepts the Born rule.

Another way to look at this is that Branch Independence forbids decisions in one branch from taking into consideration knowledge of the existence of others.
If the agent would normally consider $A\succ \mathds{1}$, his opinion can not be changed by knowledge that he is in a down branch of $B$ and there is already a set of \$1,000 branches. So he can not think that if he plays $A$ the new \$1,000 branches will simply blend into the set of preexisting ones, without increasing its size or relevance (which may be ill defined) to make up for the branches where he will pay \$100. But, again, without branch counting or the Born rule, nothing in EQM seems to imply that this would be an unreasonable line of thought.
 
\begin{ThmName}[Branching Indifference]\label{ax:BrIndif} 
Let $r\in\R$, $M\in\M$ with $M\subset r$, $\psi\in M$, and $U\in\U_M$.
If $U\psi\in r$ then $U\sim^\psi \mathds{1}_M$.
\end{ThmName}

This means the agent is indifferent to acts keeping him in the same reward subspace, reflecting the characterization of these as being determined by his preferences. Despite its name, the axiom does not assume $U$ is a branching act, and in Wallace's proof it is also used with erasures. 
Formally, such use relies on conditions of the form $\OO_{U\lvert_{M_i}}\subset r_i$ included in \nameref{ax:BrAv} and \nameref{ax:Eras}. But, as previously noted, to justify these conditions one must explain why the agent should not care about the actual processes involved in branching acts and erasures. 

For erasures, Wallace says the agent is indifferent by construction \cite[p.167]{Wallace2012}. 
Perhaps he means that only details the agent does not care about need to be erased, as states in the same reward are already equivalent in whatever matters to him. But this does not mean he must be indifferent to the erasure procedure in itself, as it may take a lot of work and time to erase all records from the environment (if it is even possible), and even his memories of the bet need to be wiped somehow. 

For branching acts, he says 
``an agent doesn't care about branching per se: if a certain operation leaves his future selves in $N$ different macrostates but doesn't change any of their rewards, he is indifferent as to whether or not the operation is performed'' \cite[p.170]{Wallace2012}. His use of the term reward here is ambiguous:
\begin{itemize}
\item if it refers to a prize, like cash, the quantum measurement performed in a branching act does not change it. But this does not mean the agent remains in the same reward subspace, which is characterized in terms of preferences and not prizes; 
\item if it refers to reward subspaces, the argument is circular: the agent's indifference is justified by claiming the branching act does not change the reward subspaces, but an act only keeps him in the same reward if he is indifferent to it.
\end{itemize}

Wallace also says that ``a  preference order which is not indifferent to branching per se would in practice be impossible to act on: branching is uncontrollable and ever-present in an Everettian universe'' \cite[p.170]{Wallace2012}.
This seems at odds with his view that coarse graining keeps such wild proliferation of branches under control. Anyway,  can we be sure an Everettian universe would be sufficiently well behaved to allow agents (if they can even exist) to make reasonable decisions and act accordingly\footnote{Of course, this brings the whole decision theoretic approach into question.}? As discussed, Wallace's argument that rationality is possible because there is a strategy satisfying his axioms is not valid, until we conclude the axioms are reasonable.

But why would the agent care about branchings? 
Maybe he thinks they alter the ratio of branches with each payoff. Wallace dismisses this since there is no well defined number of branches, as seen in section \ref{sec:Probability_Problem}.
But failure of a crude idea of branch counting does not exhaust all possibilities.
There might be more sophisticated ways to compare the size or relevance of sets of branches (there is the Born rule, but we can not assume it is physically meaningful in an Everettian universe).
In fact, as seen, \nameref{ax:DiacCons} implicitly assumes the agent believes a measure of relevance exists, even if he does not know which (again, to avoid circularity, we can not presume he believes in the Born rule).
So the question is whether he has a reason to think such unkown measure is preserved by branchings. 
We throw a few ideas, just to show he might actually think otherwise:

\begin{itemize}
\item there might be an optimal interval of coarse graining, fine enough not to mix macroscopically distinct states, yet coarse enough to reduce interference and keep the number of branches reasonably stable. The average branch number in this interval (perhaps giving smaller weights to the less stable numbers obtained with finer grainings) might be a good measure, not preserved by branchings;

\item branches with tiny Born weights can suffer much interference, so one might count only the more stable causal histories \cite{Mandolesi2017};

\item even if one can not attribute a precise number to sets of branches, obtaining a reasonably stable ratio between them is enough to estimate relative relevance;

\item there may be a qualitative way to compare the complexity of branching structures, showing that quantum measurements make them more ramified. 
\end{itemize} 
Even if these ideas fail, the point is we can not say a branching sensitive measure is impossible just because we do not know any. Not proving the relevance of branchings is not the same as proving their irrelevance. So it would be unreasonable for the agent to dismiss beforehand the possibility of such a measure, and decide as if he knew branchings could be ignored. 

This axiom has been criticized by other authors \cite{Dizadji-Bahmani2013,Dizadji-Bahmani2013a,Lewis2016,Mallah2008}. Wilson \cite{Wilson2013} criticizes Wallace's arguments, but defends the axiom on the basis of indexicalism.

\begin{ThmName}[Solution Continuity]\label{ax:SolCont}
Let $M\in\M$, $\psi\in M$ and $U,V\in\U_M$. If $U\succ^\psi V$ then $U'\succ^\psi V'$ for any $U',V'\in\U_M$ sufficiently close (in the operator norm) to $U$ and $V$.
\end{ThmName}

Wallace justifies this, and \nameref{ax:PrCont}, ``in terms of the limitations of any physically realisable agent. Any discontinuous preference order would require an agent to make arbitrarily precise distinctions between different acts, something which is not physically possible. Any preference order which could not be extended to allow for arbitrarily small changes in the acts considered would have the same requirement'' \cite[p.170]{Wallace2012}. 

These arguments seem to rely on the notion, from our physical experience, that very close acts differ only on unnoticeable, hence irrelevant, details (e.g. the value of a quantity being slightly off). But in an Everettian universe, with branches defined via Non-Probabilistic Decoherent Histories, small perturbations can create completely different new worlds. Without a Born rule, these can not be considered any less relevant, despite their small Born weights.

But let us assume the agent is really unable to distinguish close acts, so that, as Wallace argues, he can not have a discontinuous preference. This does not imply he has a (non-trivial) continuous preference.
As we will argue, it might actually lead to a breakdown in his decision-making capability (in which case we must even question the possibility of intelligent life under Everettian conditions).

First let us examine how physical imprecision affects Wallace's formalism of acts. If an act $U$, used to represent a bet, can never be executed with absolute precision, and can not be perfectly distinguished from other acts in a small neighborhood $\mathcal{N}_U$, then the bet should actually be represented by $\mathcal{N}_U$. To decide the agent should compare such neighborhoods, not individual acts in them. Of course, if preferences are continuous then it is valid to take $U$ as a representative of $\mathcal{N}_U$. Hence, in adopting a formalism of individual acts, Wallace is already assuming Solution Continuity. So let us take a step back and consider preferences between neighborhoods.

Let $U$ be an act in which an electron is prepared in a state of spin up and measured, with the agent receiving \$1,000 if the result is up, and paying \$1,000,000 if it is down. In Wallace's idealization of individual acts, we should have $U\succ \mathds{1}$.
But in reality, when offered such bet, the agent ought to compare small neighborhoods $\mathcal{N}_U$ and $\mathcal{N}_\mathds{1}$. 
Let $U_\epsilon\in\mathcal{N}_U$ be an act similar to $U$, but in which the preparation or measurement were slightly imprecise, generating branches, of total Born weight $\epsilon$, with result down. With the Born rule we could still say $U_\epsilon\succ \mathds{1}$, as for $\epsilon$ small enough such branches are negligible, and generalize the argument to conclude that $\mathcal{N}_U\succ\mathcal{N}_\mathds{1}$, in agreement with the preference obtained through the idealization.

But until we have such rule we can not equate tiny Born weights with irrelevance. So the agent might consider that any $U_\epsilon$, with $\epsilon>0$, results in a set of branches where he gets \$1,000 and another where he pays \$1,000,000, with no meaningful way to say one is larger or more relevant than the other. And he might conclude he has too much to lose if the act executed turns out to be $U_\epsilon$ instead of $U$. As in $\mathcal{N}_U$ there are infinitely many acts like $U_\epsilon$ and only one $U$, he might decide that $\mathcal{N}_U\prec\mathcal{N}_\mathds{1}$.
Hence comparison of individual acts might not be a harmless idealization,  leading to wrong results.

\nameref{ax:PrCont} makes this even worse. Small perturbations of any act can introduce all sorts of branches in the final state. When offered $U$, the agent would have to consider not only perturbations creating branches where he loses \$1,000,000, but also branches where he dies, becomes a king, gets an unicorn\footnote{There is nothing unphysical in a horned horse, even if in our world no such species has evolved.}, or anything he can imagine. Such states do exist somewhere in the Hilbert space, and \nameref{ax:PrCont} provides acts creating tiny components in them (even if, by the Born rule, such branches would be nearly impossible in our universe). 
As both $\mathcal{N}_U$ and $\mathcal{N}_\mathds{1}$ have acts creating all possible branches, comparing them becomes an impossibly complex task, and the agent might just give up deciding anything. Or decide all options are always equally preferable, since all acts, when perturbed, can lead to the same results (all possible ones), and (unless the Born rule can still be proven) he has no reason to believe some are less relevant than others. 

\begin{ThmName}[Act Nondegeneracy]\label{ax:ActNDeg}
There are $M\in\M$, $\psi\in M$ and $U,V\in\U_M$ such that $U\succ^\psi V$.
\end{ThmName}

This is not one of Wallace's axioms, but, as shown in \cite{Mandolesi2017a}, it must be included to ensure Born weights really play a role in his result. If the agent is always indifferent to all acts, the result is obtained trivially by setting equal utilities for all rewards. 
But then it would be valid even if the expected utilities had been defined with any other weights instead of the Born ones.
So in this case Wallace's result would not confer the Born weights any special significance.

A justification for this axiom might be that, in real life, decisions are seldom trivial, and strict preferences are usually present. But this is based on experience in our universe. 
As discussed above, it is conceivable that, under Everettian  conditions, the only reasonable attitude might be indifference to all acts, in all situations. 

\begin{ThmName}[Macrostate Indifference]\label{ax:MacIndif}
Let $M,M', M_1,M_2\in\M$, $\psi\in M$, $\psi'\in M'$, $U, V\in\U_{M}$, $U', V'\in\U_{M'}$, and $r_1,r_2\in\R$. If $\OO_{U},\OO_{U'}\subset M_1\wedge r_1$ and $\OO_{V},\OO_{V'}\subset M_2\wedge r_2$, then $U\succcurlyeq^{\psi} V \Leftrightarrow U'\succcurlyeq^{\psi'} V'$.
\end{ThmName}

In \cite{Mandolesi2017a}, this axiom has been shown to have many problems, which point to ambiguities in the concepts of macrostate and reward. But it has also been shown to be unnecessary for the proof, so we will not discuss it any longer.

\section{Conclusion}\label{sec:Conclusion}

In \cite{Mandolesi2017a}, Wallace's formal proof was found to be, for the most part, correct, even if some details needed fixing. But the interpretation of his result, as meaning that rational Everettian agents decide as if the Born rule were valid, is not immediate.
It demands, in particular, that his axioms can really be seen as mandates of rationality under Everettian conditions.

Unfortunately, we have found most axioms to be far from reasonable. Moreover, some concepts he uses present ambiguities and contradictions which seem hard to resolve. 
The main problems identified are:
\begin{enumerate}
\item Without the Born rule, Wallace's solution of the preferred basis problem does not work as expected. Branches might be nothing like our world, lacking complex structures or behaving erratically. This compromises the whole decision theoretic approach, which depends on narratives where agents exist, their actions have the expected consequences, and rationality is possible. 

\item Macrostates are expected to emerge from that solution, but Wallace does not detail how they are defined or what properties they are supposed to have. He gives vague ideas, which seem contradicted by some of his examples and the formalization of some axioms. 

\item It is not even clear whether macrostates describe the macroscopic state of just the system of interest, or also its environment. To justify \nameref{ax:Irrev}, history records in the environment must be separated in distinct macrostates, as the \nameref{Branching-Consistency Theorem} does not extend to history algebras.

\item The Boolean condition on events requires that branches be mutually orthogonal. This is a harmless idealization in the usual decoherent histories formalism, but in its non-probabilistic version such branches become erratic. More realistic branch decompositions are almost orthogonal, but even tiny deviations from orthogonality can have important consequences, such as branch decompositions being non-unique, and not even similar.

\item The concept of reward subspaces is ambiguous. Wallace characterizes them in terms of anything that might affect the agent's preferences. But in his justification of \nameref{ax:ReAv}, and when he states that branching acts and erasures do not change rewards, he seems to rely implicitly on a more concrete idea of rewards, as representing tangible prizes like cash. 

\item The definition of rewards as subspaces embeds an unjustified assumption about the agent's preferences: that if he is indifferent between two states in a reward, he must also be indifferent between any of them and a superposition of both.

\item \nameref{ax:ReAv} leads to a contradiction when applied in a typical state (with components in all macrostates), unless $\U_\HH=\emptyset$, which does not seem to be the case (even if it was, this seems more like a loophole than a valid condition).

\item \nameref{ax:ReAv}, \nameref{ax:BrAv} and \nameref{ax:Eras} require acts with images confined to reward subspaces. Due to physical imprecision these can hardly be obtained in practice, and as idealizations can lead to wrong results.

\item \nameref{ax:Irrev} and \nameref{ax:Eras} rely on conflicting claims about the preservation or elimination of history records.

\item Use of \nameref{ax:Eras} at several pairs of branches may require adjustments to make the acts compatible, perturbing the equality of final states. On the other hand, the idea that acts on different branches can always be made compatible leads to a contradiction when applied to a pair of erasures.

\item For \nameref{ax:StaSup}, Wallace does not explain the irrelevance of initial states, and for the acts his arguments are unconvincing. Claiming the final state embeds information about the whole process is not valid with erasures (which is how the axiom is used in the proof) or with Non-Probabilistic Decoherent Histories.

\item Wallace justifies the temporal aspect of \nameref{ax:DiacCons} (which is, anyhow, implied by \nameref{ax:StaSup}), but not why preferences in one branch can not take into account knowledge of others.
This requires the agent to believe in a (possibly unknown) way to measure or compare the relevance of sets of branches, but there is no reason why such belief should be mandated by EQM.

\item Despite its name, or Wallace's informal description and justification for it, the formal statement of \nameref{ax:BrIndif} makes no reference to branching acts, and in the proof it is also used with erasures. 

\item The reason for indifference to erasures is unclear. Even if the agent does not care about what is erased, it does not mean he is indifferent to the erasure process.

\item Wallace's arguments for indifference to branchings rely on the ambiguity of the term reward, and on assumptions we can not be sure to be valid in an Everettian universe: the possibility of acting rationally, and the impossibility of alternatives to branch counting besides the Born rule. 

\item With Non-Probabilistic Decoherent Histories, branch discontinuities compromise \nameref{ax:SolCont}, \nameref{ax:PrCont}, and a new axiom of \nameref{ax:ActNDeg} (needed to ensure Born weights play a role in Wallace's result). 

\item \nameref{ax:MacIndif}, as shown in \cite{Mandolesi2017a}, has many problems, but is unnecessary.
\end{enumerate}

As can be seen, Wallace's formalization of Deutsch's ideas, instead of confirming them, actually exposes many unforeseen difficulties.
Solving all of them, even if one believes it to be feasible (we do not), does not seem like an easy task. So one musk ask whether Wallace's result is worth the trouble. Assuming it can still be obtained from a better set of axioms, in a following article we discuss whether it can really be interpreted as showing the Born rule has a real significance in Everettian Quantum Mechanics.

\bibliographystyle{amsalpha}

\bibliography{Bibliografia_Wallace}

\end{document}